\documentclass[prd,amsmath,amssymb,superscriptaddress,floatfix,nofootinbib,reprint,10pt]{revtex4}
\usepackage{newtxtext,newtxmath}
\usepackage{amssymb,amsbsy,amsmath,amsfonts}
\usepackage{graphicx}
\usepackage{float}
\usepackage{color}
\usepackage{morefloats}
\usepackage{rotating}
\usepackage{srcltx}
\usepackage{slashed}
\usepackage{multirow}
\usepackage{verbatim}
\usepackage{hyperref}
\usepackage[dvipsnames,x11names]{xcolor}
\hypersetup{colorlinks=true,citecolor=Green}
\usepackage{tabularx}
\usepackage{bm}
\allowdisplaybreaks[4]

\usepackage{bbding}
\usepackage{threeparttable}
\usepackage[normalem]{ulem}

\usepackage[a4paper]{geometry}
\DeclareUnicodeCharacter{2212}{\ensuremath{-}}

\newcommand{\PreserveBackslash}[1]{\let\temp=\\#1\let\\=\temp}
\newcolumntype{C}[1]{>{\PreserveBackslash\centering}p{#1}}
\newcolumntype{R}[1]{>{\PreserveBackslash\raggedleft}p{#1}}
\newcolumntype{L}[1]{>{\PreserveBackslash\raggedright}p{#1}}

\begin{document}

\title{\boldmath Testing the nature of the $\Sigma^*(1430)$}

\author{Jia-Xin Lin}
\email[]{linjx@seu.edu.cn}
\affiliation{School of Physics, Southeast University, Nanjing 210094, China}%
\affiliation{Departamento de Física Teórica and IFIC, Centro Mixto Universidad de Valencia-CSIC Institutos de Investigación de Paterna, 46071 Valencia, Spain}

\author{Jing Song}
\email[]{Song-Jing@buaa.edu.cn}
\affiliation{School of Physics, Beihang University, Beijing, 102206, China}%
\affiliation{Departamento de Física Teórica and IFIC, Centro Mixto Universidad de Valencia-CSIC Institutos de Investigación de Paterna, 46071 Valencia, Spain}

\author{Miguel Albaladejo}
\email[]{Miguel.Albaladejo@ific.uv.es}
\affiliation{Instituto de F\'{\i}sica Corpuscular, Centro Mixto Universidad de Valencia-CSIC, Institutos de Investigaci\'on de Paterna, Aptdo. 22085, E-46071 Valencia, Spain}%

\author{Albert Feijoo}
\email[]{edfeijoo@ific.uv.es}
\affiliation{Instituto de F\'{\i}sica Corpuscular, Centro Mixto Universidad de Valencia-CSIC, Institutos de Investigaci\'on de Paterna, Aptdo. 22085, E-46071 Valencia, Spain}%

\author{ Eulogio Oset}
\email[]{oset@ific.uv.es}
\affiliation{Departamento de Física Teórica and IFIC, Centro Mixto Universidad de Valencia-CSIC Institutos de Investigación de Paterna, 46071 Valencia, Spain}
\affiliation{Department of Physics, Guangxi Normal University, Guilin 541004, China}

\begin{abstract}
We study the feasibility of having the $\Sigma^*(1430)$ state, predicted within the chiral unitary approach and recently reported by the Belle Collaboration, as corresponding to a state of non-molecular nature. Starting from this assumption, since the state is observed in the $\pi \Lambda$ channel, we allow the coupling to this state and relate the coupling to $\bar K N$ and $\pi \Sigma$ using $SU(3)$ symmetry arguments. We find that it is possible to have such a state with negligible coupling to the molecular components with a bare mass of the state very close to the physical mass of the $\Sigma^*(1430)$. Yet, this has consequences on other observables, since the width obtained is extremely small and incompatible with the Belle observations, and it leads to abnormally large values of the $\bar K N$ effective range. Conversely, such mismatches do not appear for large values of the bare mass of the state, but in this case we observe that the state develops large molecular components. While one can rule out a largely non-molecular nature for this state, its properties and detailed nature will need further experimental developments which one can anticipate will be coming in the near future.

\end{abstract}

\maketitle

\section{Introduction}

The baryon resonances have long attracted significant interest in hadron physics concerning their structure and nature \cite{Klempt:2009pi,Crede:2013kia}.
One of the key issues is the nature of the low-lying excited baryons with spin-parity quantum numbers $J^P = 1/2^-$.
It is well known that the $\Lambda(1405)$ is difficult to describe as a $p$-wave excited state within the framework of conventional quark models (CQM) \cite{Isgur:1978xj}.
The nature of the $\Lambda(1405)$ has been widely discussed \cite{Hyodo:2011ur}.
The original suggestion in Refs.~\cite{Dalitz:1959dn,Dalitz:1960du} that it was a $\bar{K}N$ bound state found its later support within the chiral unitary approach in coupled channels \cite{Kaiser:1995eg,Oset:1997it,Oller:2000fj,Jido:2003cb}.
Other works have proposed a hybrid structure, as containing both $qqq$ and $qqqq\bar{q}$ components \cite{Helminen:2000jb}. Recently, the discussion on the lowest-lying $1/2^-$ and $3/2^-$ $\Lambda^*$ resonances in Ref.~\cite{Nieves:2024dcz}, with special attention paid to the interplay between CQM and chiral meson-baryon molecular constituents, concluded that the two-pole pattern found for the $\Lambda(1405)$ state is a consequence of the decisive role of the $\bar{K}N$ channel in the dynamics together with the scarce influence of the $qqq$ component. Additionally, in Ref.~\cite{Conde-Correa:2024qzh}, the analysis of the $\bar{K}N$ system in the CQM framework recovers the two-pole nature of the $\Lambda(1405)$ only when other coupled meson-baryon channels are effectively taken into account via an optical potential.
On the other hand, the quark model also struggles to explain why the mass of the $N^*(1535)$ with $J^P = 1/2^-$ is higher than that of the $N^*(1440)$ with $J^P = 1/2^+$ \cite{Capstick:2000qj}.
It has been suggested that the $N^*(1535)$ may have a more complex structure than the standard three-quark configuration \cite{Wang:2023snv}. Some studies interpret it as a molecular state \cite{Kaiser:1995eg,Kaiser:1996js,Nieves:2001wt,Inoue:2001ip}, while others consider it as a mixture of three quarks and a pentaquark configuration \cite{Hannelius:2000gu,Zou:2005xy}.
Furthermore, the low-lying excited baryons $\Sigma^*$ with $J^P = 1/2^-$ remain elusive, with both their experimental existence and detailed theoretical description still not firmly established \cite{ParticleDataGroup:2024cfk}.
Studying this topic is crucial for understanding the nature of low-lying excited baryons.
The chiral unitary approach for pseudoscalar-baryon interaction that predicted two $\Lambda(1405)$ states \cite{Oller:2000fj,Jido:2003cb}, also predicts a $\Sigma^*$ state around $1430\,\rm MeV$, which has been recently observed experimentally by the Belle collaboration in Ref.~\cite{Belle:2022ywa}.
A recent review on the $\Sigma^*(1430)$ with $J^P = 1/2^-$ can be found in Ref.~\cite{Wang:2024jyk}.
In the present work, we focus on the controversial $\Sigma^*(1430)$ state.

The $\Sigma^*(1430)$ with $J^P = 1/2^-$, strangeness $S = -1$, and isospin $I=1$ around $1430\, \rm MeV$ has attracted the attention of the low-lying baryons community.
In Ref.~\cite{Oller:2000fj} it was found as a bound state and in Ref.~\cite{Jido:2003cb} as a cusp structure.
This state has also been reported in other works using the chiral unitary approach \cite{Garcia-Recio:2002yxy,Oller:2006jw,Kamiya:2016jqc,Khemchandani:2018amu}.
This state was further studied in Ref.~\cite{Roca:2013cca}, in relation to the experimental photoproduction data of the $\gamma p \to K^+ \pi^\pm \Sigma^\mp$ decay, where it appeared as a cusp within the same theoretical framework.
Moreover, it was found to manifest as a strong cusp in the processes of $\chi_{c0}(1P) \to \bar{\Sigma} \Sigma \pi$ \cite{Wang:2015qta} and $\chi_{c0}(1P) \to \bar{\Lambda} \Sigma \pi$ \cite{Liu:2017hdx} with a mass around $1430\,\rm MeV$.
In addition, the production of $\Sigma^*(1430)$ was evaluated in the decay of $\Lambda_c^+ \to \pi^+\pi^0\pi^-\Sigma^+$ using a triangle diagram in Ref.~\cite{Xie:2018gbi}.
The same study further suggested searching for the $\Sigma^*(1430)$ in the $\Lambda_c^+ \to \Lambda \pi^+ \pi^+ \pi^-$ reaction, the decay investigated in Ref.~\cite{Belle:2022ywa}, where the state was observed experimentally.

In the Belle experiment \cite{Belle:2022ywa}, the mass distributions of $\Lambda \pi^+$ and $\Lambda \pi^-$ in the $\Lambda_c^+ \to \Lambda \pi^+ \pi^+ \pi^-$ reaction were reported, revealing a clear enhancement near the $\bar{K}N$ threshold. The corresponding analysis employed two different functionals to mimic the $\Sigma^{\ast}(1430)$ state present in the $\Lambda\pi^\pm$ invariant mass distributions depending on the assumed nature of such a state. These two functionals consisted of a nonrelativistic Breit-Wigner (BW) and a Dalitz Model (DM), which effectively allows one to take into account the $\bar{K}N$ rescattering. The BW functional would correspond to a resonant state, while the DM one would describe a $\bar{K}N$ cusp. After convoluting the resolution, found to be 4\% when comparing simulations to raw data, the results of the $\Sigma^{\ast}(1430)$ mass and width for the BW scenario were $M_{\Lambda\pi^+}=1434.3\pm 0.6\pm 2.5$~MeV ($M_{\Lambda\pi^-}=1438.5\pm 0.9\pm 0.9$~MeV) and $\Gamma_{\Lambda\pi^+}=11.5\pm 2.8\pm 5.3$~MeV ($\Gamma_{\Lambda\pi^-}=33.0\pm 7.5\pm 23.6$~MeV), and the scattering lengths $a_{\bar{K}^0p}=(0.48\pm 0.32 \pm 0.38)-i(1.22\pm 0.83\pm 2.54)$~fm ($a_{K^-n}=(1.24\pm 0.57 \pm 1.56)-i(0.18\pm 0.13\pm 0.20)$~fm) for the DM scenario.\footnote{Although the $\text{Re}[a_{\bar{K}N}]$ and $\text{Im}[a_{\bar{K}N}]$ parts of the scattering lengths are reported as positive values in Ref.~\,\cite{Belle:2022ywa}, the seminal work of Ref.~\cite{Dalitz:1982tb} used in Eq. (2) of Ref.\,\cite{Belle:2022ywa} assumes a relative minus sign between the real and imaginary parts of the scattering lengths. We have taken into account this change of sign when quoting the scattering length values of Ref.\,\cite{Belle:2022ywa}.} This observation represents the first clear evidence for the existence of the $\Sigma^*(1430)$. This state, described within a molecular picture, has been recently revisited in the framework of the chiral unitary approach in Ref.~\cite{Li:2024tvo}. In that work, the $\bar{K}^0 p, \pi^+ \Sigma^0, \pi^0 \Sigma^+, \pi^+ \Lambda$, and $\eta \Sigma^+$ channels were considered, and a pole was found at $(1431.83 - i 104.75)\, \rm MeV$. This pole is slightly bound in the $\bar{K}^0 p$ channel, with a binding energy of about $4\,\rm MeV$. In that work, the correlation functions for the $\bar{K}N$ and coupled channels were evaluated and the inverse problem was faced. It was shown, assuming that these correlation functions were measured, that one can deduce from them the existence of the pole and scattering parameters for the different channels with relative precision. It is worth mentioning that the authors found a value for the $\bar{K}^0p$ scattering length of $a_{\bar{K}^0p}=0.452-i\,1.125\,\text{fm}$ using the chiral unitary approach (Table~III in Ref.~\cite{Li:2024tvo}) showing a great compatibility with the corresponding Belle output. It was also shown that the knowledge of the binding energy, the scattering length, the effective range, and the correlation functions for the relevant channels would provide very valuable information regarding the molecular compositeness of the $\Sigma^*(1430)$.

In the present work, we aim to investigate whether there exists a striking difference between the molecular picture and the compact pentaquark picture of the $\Sigma^*(1430)$. For this purpose, we start from a genuine compact (non-molecular) state, which can be observed in the $\bar{K}^0 p, \pi^+\Sigma^0, \pi^0\Sigma^+$, and $\pi^+\Lambda$ components. We assume that this genuine state corresponds to a resonance with a mass of $1432\, \rm MeV$. We then present the results for scattering length and effective range of the $\bar{K}^0 p$ channel, finding striking differences between the molecular and genuine pictures of the state. This issue has also been discussed for the $X(3872)$ and $T_{cc}(3875)^+$ cases in Refs.~\cite{Dai:2023kwv,Song:2023pdq}.

\section{Formalism}
\label{sec:forma}

\begin{figure}[t]
\begin{center}
\includegraphics[width=0.4\textwidth]{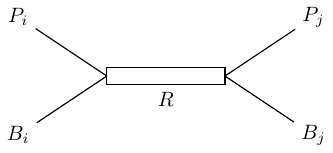}
\end{center}
\vspace{-0.7cm}
\caption{$PB$ amplitude from the genuine resonance $R$. ($P$ and $B$ stand for pseudoscalar meson and baryon, respectively.)}
\label{Fig:R}
\end{figure}

Let us assume that we have a genuine resonance with a bare mass $M_R$, which is not generated from meson-baryon interactions. Nevertheless, this state couples to $\bar{K}^0 p, \pi^+\Sigma^0, \pi^0\Sigma^+$, and $\pi^+\Lambda$, since it is observed in $\pi^+\Lambda$ and this channel couples to the other ones.
We further assume that even if the $\Sigma^*(1430)$ couples weakly to these channels, the effects of this state can still be observed in the relevant observables. It is important to note that the $\Sigma^*(1430)$ appears near the $\bar{K}N$ threshold, which is the most important component in the molecular picture, while the other channels provide the decay width of the $\Sigma^*(1430)$. The genuine state $\Sigma^*(1430)$ coupling to these channels is illustrated by Fig.~\ref{Fig:R}.

To obtain the coupling of the genuine resonance to the meson-baryon channels, we will assume $SU(3)$ symmetry.
The mesons belong to the $\mathbf{8}$ representation of $SU(3)$ and also the baryons. We have the following $SU(3)$ decomposition:
\begin{equation}
    \mathbf{8} \otimes \mathbf{8} \to \mathbf{1} \oplus \mathbf{8} \oplus \mathbf{8^\prime} \oplus \mathbf{10} \oplus \mathbf{\overline{10}} \oplus \mathbf{27}.
    \label{eq:octet}
\end{equation}
As one can see, if we construct an $SU(3)$-symmetric Lagrangian based purely on $SU(3)$ symmetry considerations, we would expect the existence of one singlet and two octets of resonances, corresponding to the symmetric and antisymmetric combinations of the octet representations. In the case of meson–meson interactions, only the symmetric octet appears in the $s$-wave due to Bose statistics. However, in the case of meson–baryon interactions, where the building blocks come from two octets of different nature, both the symmetric and antisymmetric octets can appear. For more details, see Ref.~\cite{Jido:2003cb}.
Since the $\Sigma^*(1430)$ has $I = 1$, it should correspond to an octet ($\mathbf{8}$) representation, rather than the singlet ($\mathbf{1}$), assuming that the $\mathbf{10}$ and $\mathbf{27}$ representations do not play a relevant role (we do not know, for instance, any $I = 3/2$ partner state of the $\Sigma^*(1430)$).

Let us assume that the $\Sigma^*(1430)$ is a mixture of $\mathbf{8}$ and $\mathbf{8^\prime}$, with unknown weights, and the relationship is given by
\begin{equation}
    \left\lvert {\Sigma^*(1430)} \right\rangle = \alpha \left\lvert \mathbf{8} \right\rangle + \beta \left\lvert \mathbf{8^\prime} \right\rangle,
\end{equation}
with $\beta = \pm \sqrt{1 - \alpha^2}$.
The coupled channels $\bar{K}^0 p, \pi^+\Sigma^0, \pi^0\Sigma^+$ and $\pi^+\Lambda$ with $I = 1, I_3 = 1$ are considered, which are denoted as channels $1, 2, 3, 4$, respectively. The thresholds of the $\eta \Sigma^+$ and $K^+ \Xi^0$ channels are far away from the $1432\,\rm MeV$ and are thus irrelevant to our present discussion, so they can be safely neglected. With the isospin multiplets convention $(\bar{K}^0, -K^-)$, $(-\pi^+, \pi^0, \pi^-)$, and $(-\Sigma^+, \Sigma^0, \Sigma^-)$, the couplings of the $\Sigma^*(1430)$ to the different channels are written as
\begin{subequations}\label{eq:weitht}
\begin{eqnarray}
    \langle \Sigma^*(1430) | \bar{K}^0 p \rangle &=& -\alpha \sqrt{\frac{3}{10}} - \beta\sqrt{\frac{1}{6}} = A_1\,,\\ [2mm]
    \langle \Sigma^*(1430) | \pi^+ \Sigma^0 \rangle &=&  - \beta\sqrt{\frac{1}{3}} = A_2\,,\\ [2mm]
    \langle \Sigma^*(1430) | \pi^0 \Sigma^+ \rangle &=&  \beta\sqrt{\frac{1}{3}} = A_3\,,\\ [2mm]
    \langle \Sigma^*(1430) | \pi^+ \Lambda^0 \rangle &=&  - \alpha\sqrt{\frac{1}{5}} = A_4\,,
\end{eqnarray}
\end{subequations}
where the $\alpha$ and $\beta$ coefficients are associated with the $\mathbf{8}$ and $\mathbf{8^\prime}$ representations, respectively. The transition potential between these channels due to the genuine resonance is written as
\begin{subequations}\label{eq:Vij}
\begin{align}
    V_{ij} & = \frac{\tilde{g}^2}{\sqrt{s} - M_R} C_{ij}\,,\\
    C_{ij} & = A_i A_j\,,
\end{align}
\end{subequations}
where $M_R$ is the bare mass of the resonance, and $\tilde{g}$ is the bare coupling.
%
%

%
\begin{figure}[t]
\begin{center}
\includegraphics[width=0.95\textwidth]{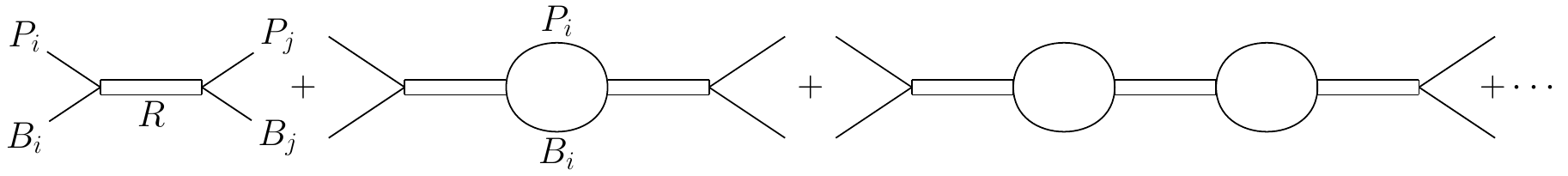}
\end{center}
\vspace{-0.7cm}
\caption{Iterated diagram of Fig.~\ref{Fig:R} implementing unitarity of the $PB$ amplitude.}
\label{Fig:iterate}
\end{figure}
Now, we unitarize the amplitude as shown in Fig.~\ref{Fig:iterate}. This procedure leads to the scattering matrix, which is given by
\begin{equation}
    T_{ij} = V_{ij} + V_{il} G_l V_{lj} + V_{il} G_l V_{lm} G_m V_{mj} + \cdots \Rightarrow
    T = V + VGT.
    \label{eq:VGT}
\end{equation}
We easily find that Eq.~\eqref{eq:VGT} becomes
\begin{equation}
    T = [1-VG]^{-1}V,
    \label{eq:T}
\end{equation}
where $G$ is the diagonal matrix whose elements, $G_i$, represent the meson-baryon loop functions, regularized using a cutoff $q_{\text{max}}$, as follows:
\begin{equation}
    G_i (\sqrt{s}) = 2M_i \int_{|\vec{q} < q_{\text{max}}|} \frac{\mathrm{d}^3 q}{(2\pi)^3} \frac{\omega_1(\vec{q}\,) + \omega_2(\vec{q}\,)}{2\omega_1(\vec{q}\,)\omega_2(\vec{q}\,)} \frac{1}{s - [\omega_1(\vec{q}\,) + \omega_2(\vec{q}\,)]^2 + i\epsilon},
    \label{eq:G}
\end{equation}
where $m_i$ and $M_i$ are the meson and baryon masses, respectively, $\omega_1(\vec{q}\,) = \sqrt{\vec{q}^{\,2} + m_i^2}$, $\omega_2(\vec{q}\,) = \sqrt{\vec{q}^{\,2} + M_i^2}$, and $q_{\rm max} = 630 \, \rm MeV $ \cite{Oset:1997it,Li:2024tvo}.
We look for possible poles of Eq.~\eqref{eq:T} in the second Riemann sheet $G_i^\text{II} (\sqrt{s})$, which is given by
\begin{equation}
    G_i^\text{II} (\sqrt{s}) = G_i(\sqrt{s}) + i \frac{M_i}{2\pi\sqrt{s}} p,
    \label{eq:G2}
\end{equation}
with
\begin{equation}
    p = \frac{\lambda^{1/2}(s, m_i^2, M_i^2)}{2\sqrt{s}},
    \label{eq:pmom}
\end{equation}
for $\mathrm{Re} (\sqrt{s}) > M_i + m_i$ and $\mathrm{Im}(p) > 0$.

The unitary matrix $T$ develops a pole at $M^\prime_R$, given by $\mathrm{det}(\mathbf{1} - VG) = 0$.
It is interesting to see the coupling of the resonance in the case of the $\bar{K}^0 p$ channel for the state, which is obtained from
\begin{equation}
    g_1^2 = \mathrm{lim} (\sqrt{s} - M_R^\prime) T_{11} |_{\sqrt{s} = M_R^\prime}.
    \label{eq:coup}
\end{equation}
The same would hold for the other channels, obtaining $g_i^2$ from $T_{ii}$, but we keep track of the relative signs by using
\begin{equation}
    g_i = \lim_{(\sqrt{s} \to M_R^\prime)} \frac{T_{1i}}{T_{11}} g_1.
    \label{eq:gi}
\end{equation}
The relationship between the scattering matrix $T$ of Eq.~\eqref{eq:T} and the quantum mechanics formalism is given by
\begin{equation}\label{eq:T-ERE-expansion}
    T = - \frac{8\pi \sqrt{s}}{2M} f^\text{QM} \simeq - \frac{8\pi \sqrt{s}}{2M} \frac{1}{\displaystyle -\frac{1}{a} + \frac{1}{2} r_0 p^2 - ip},
\end{equation}
where $a$ and $r_{0}$ are the scattering length and effective range, respectively, and $p$ is given by Eq.~\eqref{eq:pmom}. For the channel $\bar{K}p$, we can easily find:
\begin{subequations}\label{eq:ereparameters}
\begin{eqnarray}
    \label{eq:a}
    \frac{1}{a_1} &=& \left. \frac{8\pi \sqrt{s}}{2M_p} (T_{11})^{-1} \right\rvert_{\sqrt{s}_{\rm th, 1}}, \\[2mm]
    \label{eq:r}
    r_{0,1 } &=& \left. \frac{1}{\mu} \frac{\partial}{\partial\sqrt{s}} \left[ \frac{-8\pi\sqrt{s}}{2M_p}(T_{11})^{-1} + ip_1 \right] \right\rvert_{\sqrt{s}_{\rm th, 1}},
\end{eqnarray}
\end{subequations}
where $M_p$ is the mass of the proton, $\mu$ is the reduce mass of the $\bar{K}^0 p$ channel with $\mu = m_{\bar{K}^0} M_p/(m_{\bar{K}^0} + M_p)$, $\sqrt{s}_{\rm th, 1}$ is the threshold energy for the $\bar{K}^0 p$ channel, and $p_1$ is the momenta of the $\bar{K}^0,~ p$ particles in their center of the mass frame.

\section{results}
\label{sec:res}

We take the potential $V_{ij}$ from Eq.~\eqref{eq:Vij} and, for different values of $\alpha$ and $M_R$, we determine the value of $\tilde{g}^2$ that produces a pole (a zero of $\text{det}(\mathbf{1}-VG)$) at $\text{Re}\,M_R^\prime = 1432\,\text{MeV}$. The imaginary part of the pole position, $\text{Im}\,M_R^\prime$ provides the half width of the state. We then determine the scattering length and effective range from Eqs.\,\eqref{eq:ereparameters}, and the couplings of the resonance state to each coupled channel, $g_i$ from Eqs.~\eqref{eq:coup} and \eqref{eq:gi}.

\begin{table}[t]
\centering
\caption{Values of the  coupling constant $\tilde{g}^2$, half-width $\Gamma/2$, scattering length $a_1$, effective range $r_{0,1}$, and the coupling $g_i$ for various values of $\alpha$ with $\Delta M_R = 1 \, \rm MeV$ and $\text{Re}\,M_R^\prime = 1432\,\rm MeV$.}
\setlength{\tabcolsep}{7pt}
\begin{tabular}{c|cccccccc}
\hline\hline
$\alpha$ & $\tilde{g}^2$ & $\Gamma/2$\,[MeV] & $a_1$\,[fm] & $r_{0,1}$\,[fm] & $g_1$ & $g_2$ & $g_3$ & $g_4$ \\
\hline
$1.0$ & $0.62$ & $1.86$ & $0.37 -i0.17$ & $-61.74 -i0.44$ & $0.42 +i0.00$ & $0.00 +i0.00$ & $0.00 +i0.00$ & $0.35 +i0.00$ \\
$0.9$ & $0.41$ & $1.58$ & $0.38 -i0.15$ & $-62.77 -i0.46$ & $0.42 +i0.00$ & $0.16 +i0.00$ & $-0.16 -i0.00$ & $0.25 +i0.00$ \\
$0.8$ & $0.36$ & $1.66$ & $0.34 -i0.14$ & $-69.65 -i0.59$ & $0.40 +i0.00$ & $0.20 +i0.00$ & $-0.20 -i0.00$ & $0.21 +i0.00$ \\
$0.7$ & $0.33$ & $1.77$ & $0.30 -i0.13$ & $-77.44 -i0.74$ & $0.38 +i0.00$ & $0.23 +i0.00$ & $-0.23 -i0.00$ & $0.18 +i0.00$ \\
$0.6$ & $0.31$ & $1.89$ & $0.27 -i0.13$ & $-86.32 -i0.91$ & $0.36 +i0.00$ & $0.25 +i0.00$ & $-0.25 -i0.00$ & $0.15 +i0.00$ \\
$0.5$ & $0.30$ & $2.00$ & $0.24 -i0.12$ & $-96.69 -i1.11$ & $0.34 -i0.00$ & $0.27 -i0.00$ & $-0.27 +i0.00$ & $0.12 -i0.00$ \\
$0.4$ & $0.30$ & $2.12$ & $0.21 -i0.11$ & $-109.14 -i1.34$ & $0.32 -i0.00$ & $0.29 -i0.00$ & $-0.29 +i0.00$ & $0.10 -i0.00$ \\
$0.3$ & $0.30$ & $2.25$ & $0.18 -i0.10$ & $-124.58 -i1.63$ & $0.30 -i0.00$ & $0.30 -i0.00$ & $-0.30 +i0.00$ & $0.07 -i0.00$ \\
$0.2$ & $0.31$ & $2.37$ & $0.15 -i0.09$ & $-144.45 -i2.00$ & $0.28 -i0.00$ & $0.31 -i0.00$ & $-0.31 +i0.00$ & $0.05 -i0.00$ \\
$0.1$ & $0.32$ & $2.50$ & $0.12 -i0.08$ & $-171.19 -i2.51$ & $0.26 -i0.00$ & $0.32 -i0.00$ & $-0.32 +i0.00$ & $0.03 -i0.00$ \\
$0.0$ & $0.33$ & $2.63$ & $0.10 -i0.07$ & $-209.25 -i3.22$ & $0.23 -i0.00$ & $0.33 -i0.00$ & $-0.33 +i0.00$ & $0.00 +i0.00$ \\
\hline\hline
\end{tabular}
\label{Tab:res_mr1}
\end{table}
\begin{table}[t]
\centering
\caption{Same as Table~\ref{Tab:res_mr1}, but for $\Delta M_R = 10\,\text{MeV}$.%
\label{Tab:res_mr10}%
}
\setlength{\tabcolsep}{7pt}
\begin{tabular}{c|cccccccc}
\hline\hline
$\alpha$ & $\tilde{g}^2$ & $\Gamma/2$\,[MeV] & $a_1$\,[fm] & $r_{0,1}$\,[fm] & $g_1$ & $g_2$ & $g_3$ & $g_4$ \\
\hline
1.0 & 1.79 & 4.99 & $0.49 - i0.52$ & $-21.01 - i0.44$ & $0.70 + i0.01$ & $0$ & $0$ & $0.57 + i0.01$ \\
0.9 & 1.17 & 4.22 & $0.56 - i0.52$ & $-21.45 - i0.46$ & $0.69 + i0.01$ & $0.26 + i0.00$ & $-0.26 - i0.00$ & $0.42 + i0.01$ \\
0.8 & 1.02 & 4.47 & $0.48 - i0.48$ & $-23.84 - i0.59$ & $0.66 + i0.01$ & $0.34 + i0.00$ & $-0.34 - i0.00$ & $0.35 + i0.01$ \\
0.7 & 0.94 & 4.80 & $0.41 - i0.44$ & $-26.56 - i0.74$ & $0.63 + i0.01$ & $0.39 + i0.00$ & $-0.39 - i0.00$ & $0.29 + i0.00$ \\
0.6 & 0.89 & 5.14 & $0.34 - i0.40$ & $-29.65 - i0.91$ & $0.60 + i0.01$ & $0.42 + i0.00$ & $-0.42 - i0.00$ & $0.25 + i0.00$ \\
0.5 & 0.87 & 5.50 & $0.28 - i0.36$ & $-33.27 - i1.11$ & $0.57 + i0.00$ & $0.45 + i0.00$ & $-0.45 - i0.00$ & $0.20 + i0.00$ \\
0.4 & 0.86 & 5.86 & $0.23 - i0.32$ & $-37.63 - i1.34$ & $0.54 + i0.00$ & $0.48 + i0.00$ & $-0.48 - i0.00$ & $0.16 + i0.00$ \\
0.3 & 0.86 & 6.22 & $0.19 - i0.28$ & $-43.05 - i1.63$ & $0.51 - i0.00$ & $0.50 - i0.00$ & $-0.50 + i0.00$ & $0.12 - i0.00$ \\
0.2 & 0.88 & 6.60 & $0.15 - i0.24$ & $-50.02 - i2.00$ & $0.47 - i0.00$ & $0.52 - i0.00$ & $-0.52 + i0.00$ & $0.08 - i0.00$ \\
0.1 & 0.90 & 6.99 & $0.12 - i0.20$ & $-59.43 - i2.51$ & $0.44 - i0.00$ & $0.54 - i0.00$ & $-0.54 + i0.00$ & $0.04 - i0.00$ \\
0.0 & 0.94 & 7.39 & $0.09 - i0.16$ & $-72.83 - i3.22$ & $0.39 - i0.00$ & $0.56 - i0.01$ & $-0.56 + i0.01$ & $0$ \\
\hline\hline
\end{tabular}
\end{table}
\begin{table}[t]
\centering
\caption{Same as Table~\ref{Tab:res_mr1}, but for $\Delta M_R = 25\,\text{MeV}$.%
\label{Tab:res_mr25}%
}
\setlength{\tabcolsep}{7pt}
\begin{tabular}{c|cccccccc}
\hline\hline
$\alpha$ & $\tilde{g}^2$ & $\Gamma/2$\,[MeV] & $a_1$\,[fm] & $r_{0,1}$\,[fm] & $g_1$ & $g_2$ & $g_3$ & $g_4$ \\
\hline
1.0 & 3.78 & 9.84 & $0.43 - i0.71$ & $-9.52 - i0.44$ & $1.01 + i0.04$ & $0$ & $0$ & $0.82 + i0.03$ \\
0.9 & 2.46 & 8.22 & $0.53 - i0.76$ & $-9.77 - i0.46$ & $0.99 + i0.04$ & $0.37 + i0.01$ & $-0.37 - i0.01$ & $0.59 + i0.02$ \\
0.8 & 2.14 & 8.81 & $0.43 - i0.69$ & $-10.91 - i0.59$ & $0.95 + i0.03$ & $0.48 + i0.02$ & $-0.48 - i0.02$ & $0.50 + i0.02$ \\
0.7 & 1.96 & 9.54 & $0.35 - i0.62$ & $-12.19 - i0.74$ & $0.91 + i0.02$ & $0.55 + i0.01$ & $-0.55 - i0.01$ & $0.42 + i0.01$ \\
0.6 & 1.86 & 10.32 & $0.28 - i0.55$ & $-13.66 - i0.91$ & $0.87 + i0.01$ & $0.61 + i0.01$ & $-0.61 - i0.01$ & $0.36 + i0.01$ \\
0.5 & 1.81 & 11.10 & $0.22 - i0.49$ & $-15.38 - i1.11$ & $0.82 + i0.01$ & $0.66 + i0.01$ & $-0.66 - i0.01$ & $0.29 + i0.00$ \\
0.4 & 1.79 & 11.90 & $0.17 - i0.42$ & $-17.46 - i1.34$ & $0.78 + i0.00$ & $0.70 + i0.00$ & $-0.70 - i0.00$ & $0.24 + i0.00$ \\
0.3 & 1.79 & 12.71 & $0.13 - i0.36$ & $-20.05 - i1.63$ & $0.73 - i0.00$ & $0.73 - i0.00$ & $-0.73 + i0.00$ & $0.18 - i0.00$ \\
0.2 & 1.81 & 13.54 & $0.10 - i0.30$ & $-23.40 - i2.00$ & $0.68 - i0.01$ & $0.76 - i0.01$ & $-0.76 + i0.01$ & $0.12 - i0.00$ \\
0.1 & 1.86 & 14.38 & $0.07 - i0.25$ & $-27.93 - i2.51$ & $0.63 - i0.01$ & $0.79 - i0.01$ & $-0.79 + i0.01$ & $0.06 - i0.00$ \\
0.0 & 1.93 & 15.24 & $0.05 - i0.20$ & $-34.39 - i3.22$ & $0.57 - i0.01$ & $0.81 - i0.02$ & $-0.81 + i0.02$ & $0$ \\
\hline\hline
\end{tabular}
\end{table}
\begin{table}[t]
\centering
\caption{Same as Table~\ref{Tab:res_mr1}, but for $\Delta M_R = 100\,\text{MeV}$.%
\label{Tab:res_mr100}%
}
\setlength{\tabcolsep}{7pt}
\begin{tabular}{c|cccccccc}
\hline\hline
$\alpha$ & $\tilde{g}^2$ & $\Gamma/2$\,[MeV] & $a_1$\,[fm] & $r_{0,1}$\,[fm] & $g_1$ & $g_2$ & $g_3$ & $g_4$ \\
\hline
$1.00$ & $14.36$ & $32.89$ & $0.39 - i0.78$ & $-1.92 - i0.44$ & $1.99 + i0.16$ & $0$ & $0$ & $1.62 + i0.13$ \\
$0.90$ & $9.27$ & $26.29$ & $0.47 - i0.87$ & $-2.04 - i0.46$ & $1.91 + i0.16$ & $0.72 + i0.06$ & $-0.72 - i0.06$ & $1.14 + i0.10$ \\
$0.80$ & $7.96$ & $29.07$ & $0.36 - i0.80$ & $-2.33 - i0.59$ & $1.86 + i0.11$ & $0.94 + i0.06$ & $-0.94 - i0.06$ & $0.97 + i0.06$ \\
$0.70$ & $7.24$ & $32.40$ & $0.27 - i0.72$ & $-2.67 - i0.74$ & $1.80 + i0.05$ & $1.10 + i0.03$ & $-1.10 - i0.03$ & $0.83 + i0.02$ \\
$0.60$ & $6.79$ & $35.85$ & $0.19 - i0.64$ & $-3.06 - i0.91$ & $1.73 - i0.00$ & $1.22 - i0.00$ & $-1.22 + i0.00$ & $0.71 - i0.00$ \\
$0.50$ & $6.50$ & $39.29$ & $0.12 - i0.56$ & $-3.52 - i1.11$ & $1.65 - i0.04$ & $1.31 - i0.03$ & $-1.31 + i0.03$ & $0.59 - i0.01$ \\
$0.40$ & $6.33$ & $42.69$ & $0.07 - i0.48$ & $-4.09 - i1.34$ & $1.56 - i0.08$ & $1.39 - i0.07$ & $-1.39 + i0.07$ & $0.47 - i0.02$ \\
$0.30$ & $6.24$ & $46.06$ & $0.04 - i0.41$ & $-4.81 - i1.63$ & $1.46 - i0.10$ & $1.45 - i0.10$ & $-1.45 + i0.10$ & $0.35 - i0.02$ \\
$0.20$ & $6.23$ & $49.38$ & $0.01 - i0.33$ & $-5.75 - i2.00$ & $1.35 - i0.12$ & $1.50 - i0.13$ & $-1.50 + i0.13$ & $0.24 - i0.02$ \\
$0.10$ & $6.29$ & $52.66$ & $-0.01 - i0.27$ & $-7.02 - i2.51$ & $1.23 - i0.13$ & $1.54 - i0.16$ & $-1.54 + i0.16$ & $0.12 - i0.01$ \\
$0.00$ & $6.42$ & $55.92$ & $-0.02 - i0.21$ & $-8.84 - i3.22$ & $1.11 - i0.13$ & $1.57 - i0.19$ & $-1.57 + i0.19$ & $0$ \\
\hline\hline
\end{tabular}
\end{table}

At first, we show in Tables~\ref{Tab:res_mr1}, \ref{Tab:res_mr10}, \ref{Tab:res_mr25} and \ref{Tab:res_mr100} the results for the squared coupling constant $\tilde{g}^2$, half-width $\Gamma/2$, scattering length $a_1$, and effective range $r_{0,1}$ at the  threshold of the $\bar{K}p$ channel for different values of $\alpha$ and $M_R$. The resonance mass is fixed at Re$M_R^\prime = 1432\,\text{MeV}$, and $M_R$ is defined as $M_R = M_{\mathrm{th}, 1} + \Delta M_R$, where $M_{\mathrm{th},1}$ is the $\bar{K}^0p$ threshold and $\Delta M_R$ is taken to be $1$, $10$, $25$, and $100\,\text{MeV}$.

For $\Delta M_R = 1\,\rm MeV$, corresponding to a narrow near-threshold structure, the coupling $\tilde{g}^2$ decreases from $0.62$ to $0.33$ as $\alpha$ decreases from one to zero, stabilizing at small values of $\alpha$. The half-width $\Gamma/2$ increases gradually from $1.86 \,\text{MeV}$ to $2.63 \,\text{MeV}$. The real part of the scattering length $a_1$ decreases from $0.37 \,\rm fm$ to $0.10 \,\rm fm$, indicating a weakening of the interaction strength. The effective range $r_{0,1}$ becomes significantly more negative, with the real part evolving from $-61.74\,\text{fm}$ to $-209.25\,\text{fm}$. The widths obtained with this scenario are small compared with the width from the Belle experiment \cite{Belle:2022ywa} of $\Gamma_{\Lambda\pi^+}\in[5.5,17.5]\,\text{MeV}$ for the  $\Lambda\pi^+$ invariant mass distribution and $\Gamma_{\Lambda\pi^-}\in[8.2,57.8]\,\text{MeV}$ for the $\Lambda\pi^-$ one, which certainly leads to a compatibility range in $\Gamma_{\Lambda\pi}$ for the Belle results. Also, the values of $r_{0,1}$ look abnormally large, although this is not yet experimentally measured. 

As we go to $\Delta M_R=10$ and $25\,\text{MeV}$, we observe that the widths become gradually larger, the scattering lengths remain with similar values, and the effective ranges decrease, but the magnitudes of the latter still look abnormally large. The experimental determination of this magnitude, something feasible using correlation functions, would be of great help to eventually rule out this scenario. Note that the couplings $\widetilde{g}$ and $g_i$ become increasingly larger as one increases $\Delta M_R$.

In the case of $\Delta M_R = 100\,\rm MeV$, corresponding to a broader and more distant bare resonance, $\tilde{g}^2$ takes on larger values, decreasing from $14$ to $6$ as $\alpha$ decreases. The half-width $\Gamma/2$ also increases, with values ranging from $33\,\text{MeV}$ to $56\,\text{MeV}$ for increasing $\alpha$. Hence, the widths seem too large compared with the experimental one. The real part of the scattering length $a_1$ shows a decreasing trend and even becomes negative for small $\alpha$, which points to a qualitative change in the interaction. The effective range $r_{0,1}$ becomes more negative, with the real part evolving from $-1.92$\,fm to $-8.84\,\text{fm}$.

\begin{figure}[t]
\begin{center}
\includegraphics[width=0.42\textwidth]{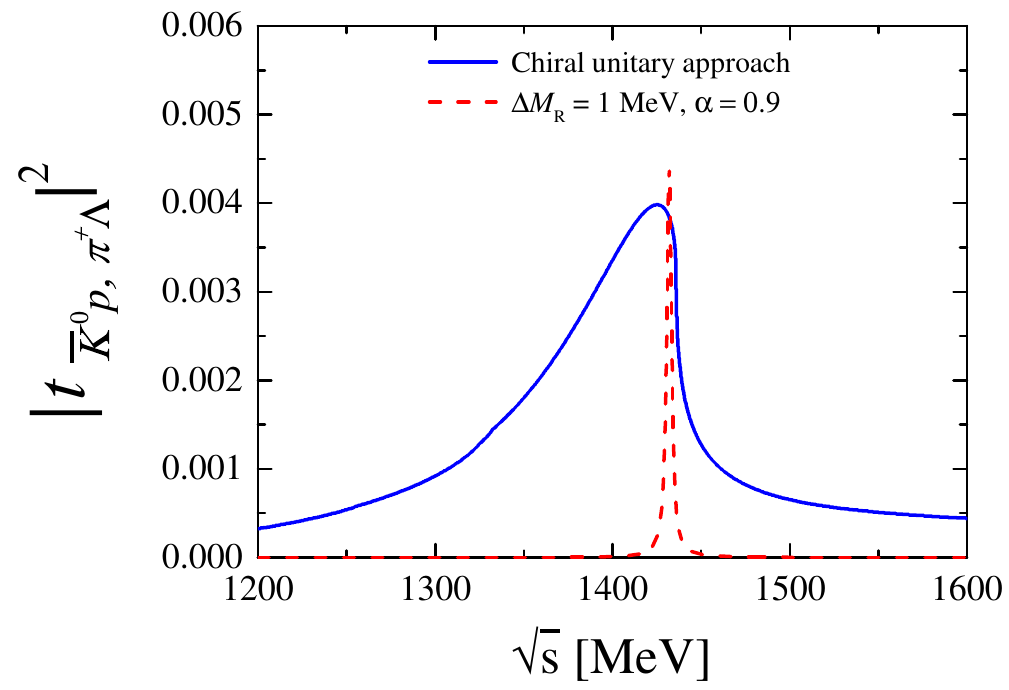}
\includegraphics[width=0.42\textwidth]{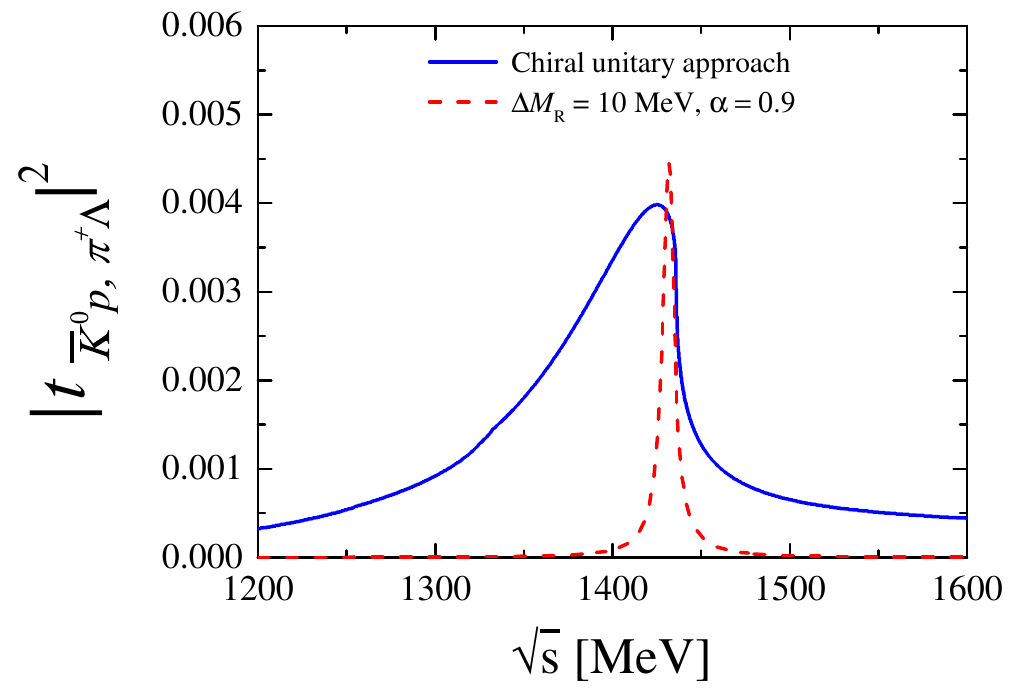}\\
\includegraphics[width=0.42\textwidth]{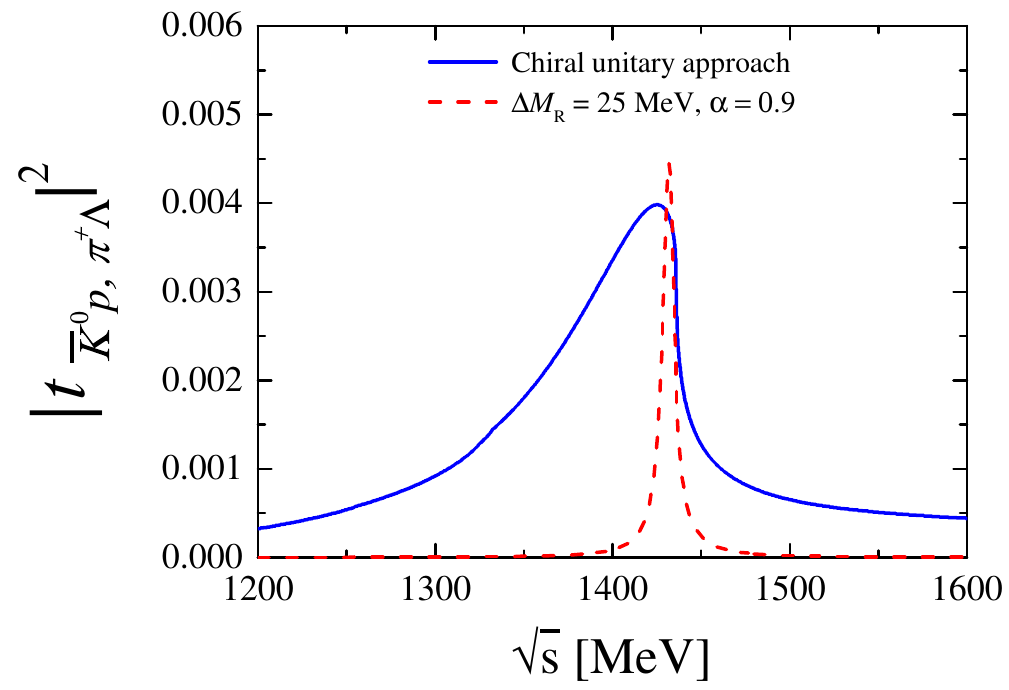}
\includegraphics[width=0.42\textwidth]{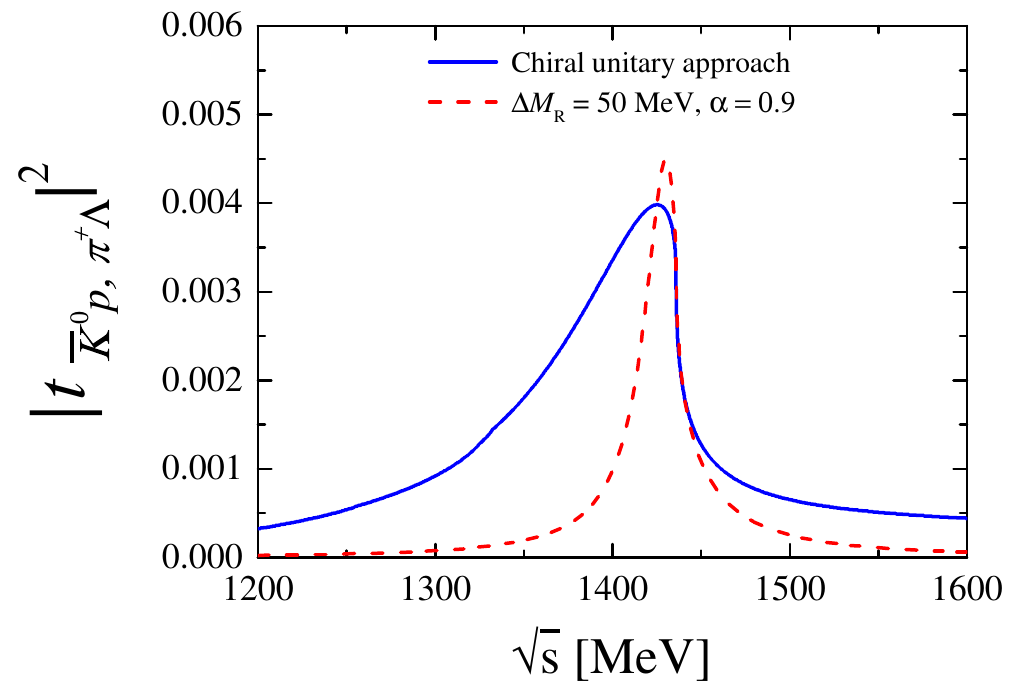}\\
\includegraphics[width=0.42\textwidth]{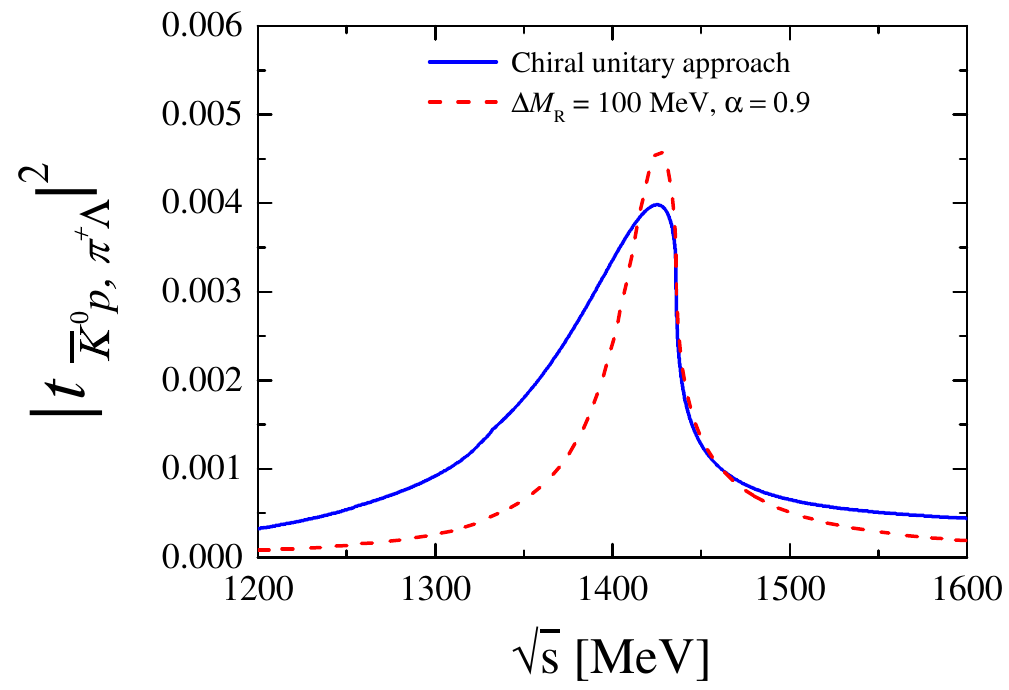}
\includegraphics[width=0.42\textwidth]{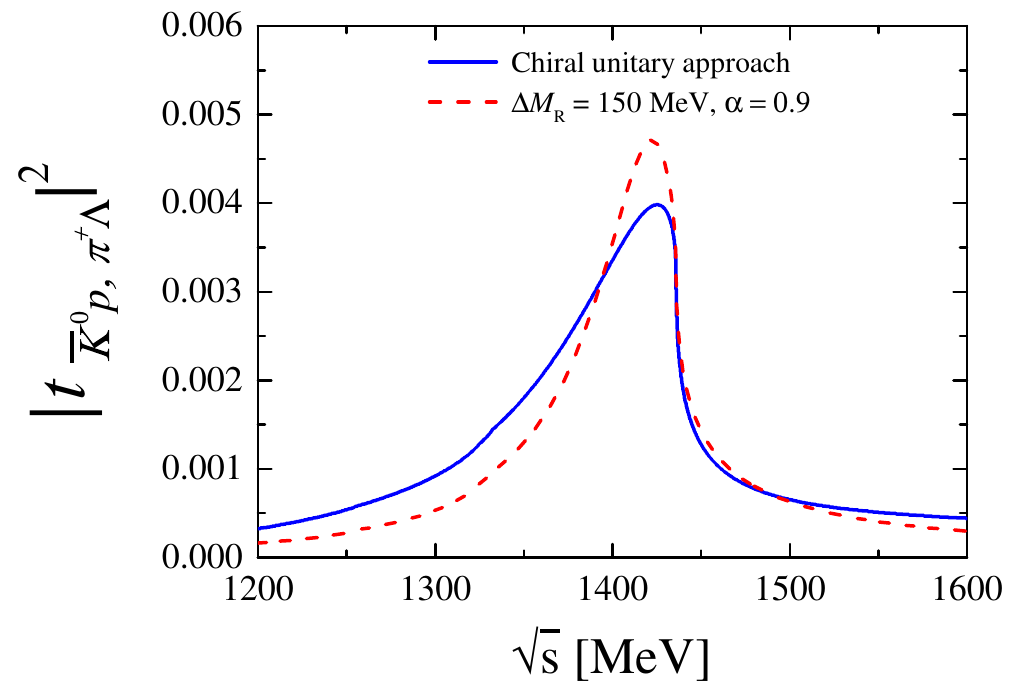}
\end{center}
\vspace{-0.7cm}
\caption{Absolute squared amplitude $|T_{\bar{K}^0 p \to \pi^+\Lambda}|^2$ as a function of the invariant mass. The blue solid line represents the full meson-baryon interaction model from Ref.~\cite{Li:2024tvo}, while the red dashed line corresponds to the model without explicit meson-baryon dynamics.%
\label{Fig:t_square}}
\end{figure}

At this point it is worth mentioning that for states close to a threshold, the imaginary part of the pole position might be misleading concerning the actual width that one observes in an experiment, as it was shown explicitly in Ref.~\cite{Li:2024tvo} where $2\,\text{Im}\,M_R'$ for the $\Sigma^*(1430)$ was much bigger than the width provided by $|T_{\bar K^0p,~\pi^+\Lambda}|^2$ around the $\bar KN$ threshold. Considering this fact, we also show in Fig.~\ref{Fig:t_square} the value of $|T_{\bar K^0p,~\pi^+\Lambda}|^2$ as a function of $\sqrt{s}$, for different values of $\Delta M_R$. 
To simplify, we choose a value of $\alpha =0.9$, in all cases, but the results are similar for other values of $\alpha$. For comparison we also show the results for this magnitude with the chiral unitary approach. 
We can see that for $\Delta M_R$ small the apparent width is indeed very small and we can dismiss this scenario. 
However, we can also see that for $\Delta M_R=100\,\text{MeV}$ the apparent width is smaller than the one of $2\,\text{Im}\,M_R'$. Hence, we should not dismiss that case either.

\begin{table}[t]
\centering
\caption{Values of the coupling and compositeness in the chiral unitary approach.}
\setlength{\tabcolsep}{25pt}
\begin{tabular}{cccccc}
\hline\hline
$g_{\bar{K}^0 p}$ & $g_{\pi^+\Sigma^0}$ & $g_{\pi^0\Sigma^+}$ & $g_{\pi^+\Lambda}$ & $g_{\eta\Sigma^+}$ \\
\hline
$3.03 - i2.71$ & $1.98 - i1.45$ & $-1.98 + i1.47$ & $0.19 - i1.21$ & $0.21 - i1.27$ \\
\hline
$P_{\bar{K}^0 p}$ & $P_{\pi^+\Sigma^0}$ & $P_{\pi^0\Sigma^+}$ & $P_{\pi^+\Lambda}$ & $P_{\eta\Sigma^+}$ \\
\hline
$-0.60 - i0.19$ & $0.24 + i0.41$ & $0.25 + i0.41$ & $0.09 - i0.05$ & $-0.01 + i0.00$ \\
\hline\hline
\end{tabular}
\label{Tab:res_chua}
\end{table}

A scenario similar to the one obtained in Table~\ref{Tab:res_mr100} with $\Delta M_R = 100$\,MeV does not seem unreasonable. For values of $\alpha \simeq 0.7-1$ the values of $\Gamma$ are reasonable and $a_1 \sim 0.4 - i 0.8$\,fm. These values would be comparable to those found in the chiral unitary approach in Ref~\cite{Li:2024tvo} $a_1 = 0.452 - i 1.125\,\text{fm}$. The effective range obtained in this scenario is $r_{0,1} \sim 2-i0.50\,\text{fm}$. This diverts somewhat from the values of Ref~\cite{Li:2024tvo}, $r_{0,1} = 0.043 - i 0.451$\,fm, although the imaginary parts are similar. We can also compare the couplings $g_i$ obtained in that scenario with those obtained within the chiral unitary approach of Ref~\cite{Li:2024tvo}, which we show in Table~\ref{Tab:res_chua}. The real part of $g_{\bar{K}^0p}$ in the scenario of Table~\ref{Tab:res_mr100} is $\text{Re}\,g_1 \sim 1.90$ for $\alpha$ values close to one, while this value is around $3$ in the chiral unitary approach. There are however more differences in $\text{Im}\,g_1$ and in the couplings to the other channels, but the values are of the same order of magnitude.

\begin{figure}[t]
\begin{center}
\includegraphics[height=5cm,keepaspectratio]{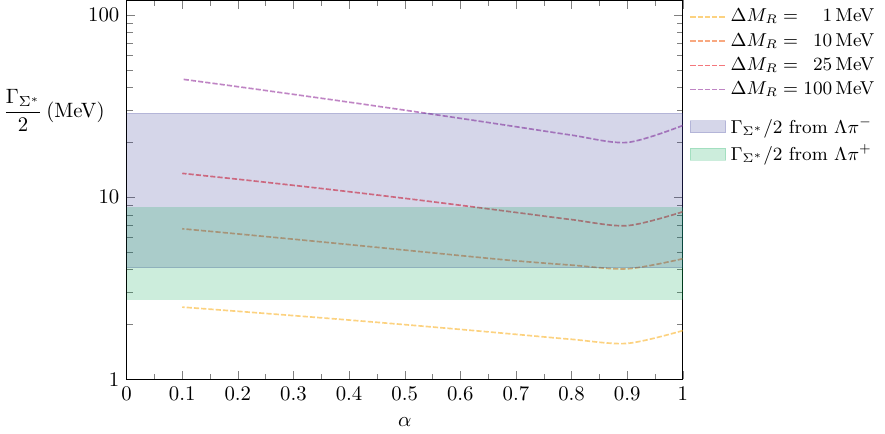}
\end{center}
\vspace{-0.7cm}
\caption{Fig: Results for $\Gamma_{\Sigma^*}/2$\,[MeV] extracted from the two experimental bands correspond to the Belle experiment \cite{Belle:2022ywa}, the cyan from the $\Lambda\pi^+$ spectrum and light gray from the $\Lambda\pi^-$ spectrum.
\label{Fig:gamma_half}}
\end{figure}

To make the discussion more visual we show in Fig.~\ref{Fig:gamma_half} the values of $\Gamma_{\Sigma^*}/2$ obtained from the width of $|T|^2$ at half value of the maximum as function of $\alpha$ for different values as $\Delta M_R$. We can see clearly that for small values of $\Delta M_R$, the results for $\Gamma_{\Sigma^*}/2$ are smaller than the experimental values, for any value of $\alpha$ particularly for large values of $\alpha$.

Intuitively, the scenario of Table~\ref{Tab:res_mr100} is not much different from the one of the chiral unitary approach, apparently indicating that one could not distinguish these pictures.
At this point we engage in a complementary discussion by looking at the compositeness of the state obtained.

The compositeness of a state in a certain channel measures the probability to find this component in the wave functions \cite{Weinberg:1965zz,Guo:2017jvc} (see also recent discussions in Ref.\,\cite{Albaladejo:2022sux}).
This is a well defined concept when one has a bound state in all the channels stemming from energy independent potentials, and these probabilities are given by \cite{Gamermann:2009uq,Hyodo:2013nka}
\begin{equation}
    P_i = -g_i^2 \frac{\partial G_i}{\partial \sqrt{s}},
    \label{eq:pprob}
\end{equation}
where $g_i$ are the couplings of Eqs.~\eqref{eq:coup} and \eqref{eq:gi}, and $G$ the loop function of Eq.~\eqref{eq:G2}. The issue becomes more involved when some of the channels are open, since the couplings and some of the $G_i$ functions become complex and Eq.~\eqref{eq:pprob} cannot be interpreted as a probability. Note that for open channels with an asymptotic wave function going as $e^{ikr/r}$, the probability becomes infinite. The issue of the compositeness of hadrons has attracted much interest~\cite{Baru:2003qq,Baru:2010ww,Baru:2021ldu,Hyodo:2011qc,Hanhart:2011jz,Aceti:2012dd,Hyodo:2013iga,Sekihara:2014kya,Hanhart:2014ssa,Guo:2015daa,Sekihara:2015gvw,Kamiya:2015aea,Albaladejo:2016hae,Sekihara:2016xnq,Kamiya:2016oao,Albaladejo:2018mhb,Matuschek:2020gqe,Albaladejo:2021cxj,Albaladejo:2021vln,Li:2021cue,Dai:2023cyo,Feijoo:2023sfe,Shi:2024llv,Kinugawa:2024crb,Montesinos:2024uhq,Kinugawa:2024kwb,Nieves:2024dcz} and different suggestions are made to still draw some meaning from the compositeness of Eq.~\eqref{eq:pprob}. Yet, the example discussed above should be sufficient to avoid associating this magnitude to probabilities.

\begin{table}[t]
\centering
\caption{Compositeness $P_i$ for channel $i$ at various values of $\alpha$, with $\Delta M_R = 1 \, \rm MeV$ and Re$M_R^\prime = 1432\,\rm MeV$.%
\label{Tab:pi1}%
}
\setlength{\tabcolsep}{22pt}
\begin{tabular}{c|cccc}
\hline\hline
$\alpha$ & $P_1$ & $P_2$ & $P_3$ & $P_4$ \\
\hline
$1.0$ & $0.05 - i0.01$ & $-0.00 + i0.00$ & $-0.00 + i0.00$ & $-0.00 + i0.01$ \\
$0.9$ & $0.05 - i0.01$ & $-0.00 + i0.00$ & $-0.00 + i0.00$ & $-0.00 + i0.00$ \\
$0.8$ & $0.04 - i0.01$ & $-0.00 + i0.00$ & $-0.00 + i0.00$ & $-0.00 + i0.00$ \\
$0.7$ & $0.04 - i0.01$ & $-0.00 + i0.00$ & $-0.00 + i0.00$ & $-0.00 + i0.00$ \\
$0.6$ & $0.03 - i0.01$ & $-0.00 + i0.00$ & $-0.00 + i0.00$ & $-0.00 + i0.00$ \\
$0.5$ & $0.03 - i0.01$ & $-0.00 + i0.00$ & $-0.00 + i0.00$ & $-0.00 + i0.00$ \\
$0.4$ & $0.03 - i0.01$ & $-0.00 + i0.01$ & $-0.00 + i0.01$ & $-0.00 + i0.00$ \\
$0.3$ & $0.02 - i0.01$ & $-0.00 + i0.01$ & $-0.00 + i0.01$ & $-0.00 + i0.00$ \\
$0.2$ & $0.02 - i0.01$ & $-0.00 + i0.01$ & $-0.00 + i0.01$ & $-0.00 + i0.00$ \\
$0.1$ & $0.02 - i0.01$ & $-0.00 + i0.01$ & $-0.00 + i0.01$ & $-0.00 + i0.00$ \\
$0.0$ & $0.01 - i0.01$ & $-0.00 + i0.01$ & $-0.00 + i0.01$ & $-0.00 + i0.00$ \\
\hline\hline
\end{tabular}
\end{table}
\begin{table}[t]
\centering
\caption{Same as Table~\ref{Tab:pi1}, but for $\Delta M_R = 10\,\text{MeV}$.%
\label{Tab:pi10}%
}
\setlength{\tabcolsep}{22pt}
\begin{tabular}{c|cccc}
\hline\hline
$\alpha$ & $P_1$ & $P_2$ & $P_3$ & $P_4$ \\
\hline
$1.0$ & $0.10 - i0.05$ & $-0.00 + i0.00$ & $-0.00 + i0.00$ & $-0.01 + i0.01$ \\
$0.9$ & $0.10 - i0.05$ & $-0.00 + i0.00$ & $-0.00 + i0.00$ & $-0.01 + i0.01$ \\
$0.8$ & $0.09 - i0.05$ & $-0.00 + i0.01$ & $-0.00 + i0.01$ & $-0.00 + i0.01$ \\
$0.7$ & $0.08 - i0.04$ & $-0.01 + i0.01$ & $-0.01 + i0.01$ & $-0.00 + i0.00$ \\
$0.6$ & $0.07 - i0.04$ & $-0.01 + i0.01$ & $-0.01 + i0.01$ & $-0.00 + i0.00$ \\
$0.5$ & $0.06 - i0.04$ & $-0.01 + i0.01$ & $-0.01 + i0.01$ & $-0.00 + i0.00$ \\
$0.4$ & $0.05 - i0.03$ & $-0.01 + i0.01$ & $-0.01 + i0.01$ & $-0.00 + i0.00$ \\
$0.3$ & $0.04 - i0.03$ & $-0.01 + i0.02$ & $-0.01 + i0.02$ & $-0.00 + i0.00$ \\
$0.2$ & $0.04 - i0.03$ & $-0.01 + i0.02$ & $-0.01 + i0.02$ & $-0.00 + i0.00$ \\
$0.1$ & $0.03 - i0.02$ & $-0.01 + i0.02$ & $-0.01 + i0.02$ & $-0.00 + i0.00$ \\
$0.0$ & $0.02 - i0.02$ & $-0.01 + i0.02$ & $-0.01 + i0.02$ & $-0.00 + i0.00$ \\
\hline\hline
\end{tabular}
\end{table}
\begin{table}[t]
\centering
\caption{Same as Table~\ref{Tab:pi1}, but for $\Delta M_R = 25\,\text{MeV}$.%
\label{Tab:pi25}%
}
\setlength{\tabcolsep}{22pt}
\begin{tabular}{c|cccc}
\hline\hline
$\alpha$ & $P_1$ & $P_2$ & $P_3$ & $P_4$ \\
\hline
$1.0$ & $0.14 - i0.10$ & $-0.00 + i0.00$ & $-0.00 + i0.00$ & $-0.03 + i0.03$ \\
$0.9$ & $0.15 - i0.10$ & $-0.01 + i0.01$ & $-0.01 + i0.01$ & $-0.01 + i0.02$ \\
$0.8$ & $0.13 - i0.10$ & $-0.01 + i0.01$ & $-0.01 + i0.01$ & $-0.01 + i0.01$ \\
$0.7$ & $0.11 - i0.09$ & $-0.01 + i0.02$ & $-0.01 + i0.02$ & $-0.01 + i0.01$ \\
$0.6$ & $0.09 - i0.08$ & $-0.01 + i0.02$ & $-0.01 + i0.02$ & $-0.01 + i0.01$ \\
$0.5$ & $0.08 - i0.07$ & $-0.02 + i0.03$ & $-0.02 + i0.03$ & $-0.00 + i0.00$ \\
$0.4$ & $0.07 - i0.07$ & $-0.02 + i0.03$ & $-0.02 + i0.03$ & $-0.00 + i0.00$ \\
$0.3$ & $0.05 - i0.06$ & $-0.02 + i0.03$ & $-0.02 + i0.03$ & $-0.00 + i0.00$ \\
$0.2$ & $0.04 - i0.05$ & $-0.02 + i0.04$ & $-0.02 + i0.04$ & $-0.00 + i0.00$ \\
$0.1$ & $0.04 - i0.04$ & $-0.02 + i0.04$ & $-0.02 + i0.04$ & $-0.00 + i0.00$ \\
$0.0$ & $0.03 - i0.03$ & $-0.02 + i0.04$ & $-0.02 + i0.04$ & $-0.00 + i0.00$ \\
\hline\hline
\end{tabular}
\end{table}
\begin{table}[t]
\centering
\caption{Same as Table~\ref{Tab:pi1}, but for $\Delta M_R = 100\,\text{MeV}$.%
\label{Tab:pi100}%
}
\setlength{\tabcolsep}{22pt}
\begin{tabular}{c|cccc}
\hline\hline
$\alpha$ & $P_1$ & $P_2$ & $P_3$ & $P_4$ \\
\hline
$1.0$ & $0.22 - i0.26$ & $-0.00 + i0.00$ & $-0.00 + i0.00$ & $-0.13 + i0.11$ \\
$0.9$ & $0.25 - i0.26$ & $-0.03 + i0.03$ & $-0.02 + i0.03$ & $-0.06 + i0.05$ \\
$0.8$ & $0.20 - i0.24$ & $-0.04 + i0.05$ & $-0.04 + i0.05$ & $-0.04 + i0.04$ \\
$0.7$ & $0.16 - i0.23$ & $-0.05 + i0.07$ & $-0.05 + i0.07$ & $-0.03 + i0.03$ \\
$0.6$ & $0.12 - i0.21$ & $-0.06 + i0.09$ & $-0.06 + i0.09$ & $-0.02 + i0.02$ \\
$0.5$ & $0.09 - i0.19$ & $-0.07 + i0.11$ & $-0.07 + i0.11$ & $-0.01 + i0.02$ \\
$0.4$ & $0.07 - i0.16$ & $-0.07 + i0.13$ & $-0.07 + i0.13$ & $-0.01 + i0.01$ \\
$0.3$ & $0.05 - i0.14$ & $-0.07 + i0.14$ & $-0.07 + i0.14$ & $-0.00 + i0.01$ \\
$0.2$ & $0.03 - i0.12$ & $-0.07 + i0.16$ & $-0.07 + i0.15$ & $-0.00 + i0.00$ \\
$0.1$ & $0.02 - i0.09$ & $-0.07 + i0.17$ & $-0.07 + i0.17$ & $-0.00 + i0.00$ \\
$0.0$ & $0.01 - i0.07$ & $-0.07 + i0.18$ & $-0.07 + i0.17$ & $-0.00 + i0.00$ \\
\hline\hline
\end{tabular}
\end{table}

Our view on the subject is discussed in Ref.~\cite{Aceti:2014ala} (sections 5 and 6 of that reference). The meaning of Eq.~\eqref{eq:pprob} is the integral of the square of the wave function (with a certain phase prescription), not the modulus square. In this magnitude, the square of the asymptotic wave function behaves as $\frac{e^{2ikr/r}}{r^2}$, rather than $\frac{1}{r^2}$ in the modulus, and the contribution of the asymptotic region vanishes in the integral, leaving thus a contribution to the integral from the confined region of the particles in the wave function. This magnitude is still meaningful, providing an idea of the ``weight" of one component in the wave function. From this perspective we analyze the results obtained from the Tables~\ref{Tab:res_mr1}, \ref{Tab:res_mr10}, \ref{Tab:res_mr25}, and \ref{Tab:res_mr100}. We show these results in Tables~\ref{Tab:pi1}, ~\ref{Tab:pi10}, ~\ref{Tab:pi25}, and \ref{Tab:pi100}, respectively. What we find from Table~\ref{Tab:pi1} is that for $\Delta M_R = 1 \, \rm MeV$ the $P_i$ values are extremely small, practically zero for all channels except the $\bar K^0p$ channel where $P_i$ is of the order of a few percent. This extremely small weight of the molecular components indicate that we would be facing a genuine non-molecular state. However, the small width obtained and the abnormally large effective range that we found in Table~\ref{Tab:res_mr1} should be sufficient to rule out this scenario. This situation reminds one of similar conclusions obtained for the $X(3872)$ from the study of the line shape for $X(3872)$ production \cite{Ji:2025hjw}.

The case with $\Delta M_R = 100\,\text{MeV}$, shown in Table~\ref{Tab:pi100} provides, however, values of $P_i$ not small compared to unity. If we take $|P_i|$ we find values about $50\%$ for $\Delta M_R = 100 \, \rm MeV$ and $\alpha=0.7-1$, which, as shown in Fig.~\ref{Fig:gamma_half}, do not have a too large width compared with the upper range of the experimental one. These numbers differ somewhat from those obtained from the chiral approach shown in Table~\ref{Tab:res_chua}, but they are one order of magnitude larger than those obtained in Table~\ref{Tab:pi1}, \textit{i.e.}, for $\Delta M_R = 1\,\text{MeV}$. This suggests that the unavoidable coupling of the original bare, non-molecular state, to the meson-baryon components dresses up the original bare state with a large molecular cloud of meson-baryon components. In simple words, the original non-molecular state has become basically molecular through the unavoidable coupling to meson-baryon components demanded by unitarity in coupled channels. This result follows the trend observed in the evolution of genuine states to molecular states in the case of $T_{cc}(3875)^+$~\cite{Dai:2023kwv} and $X(3872)$~\cite{Song:2023pdq}.

We have also checked the results for $\beta = - \sqrt{1 - \alpha^2}$. For $\alpha=1$ and $\alpha=0$ one has $\beta=0$ and $\beta=1$, respectively, and the same results as before are obtained. For $0.1 < \alpha < 0.9$ and $\beta = - \sqrt{1-\alpha^2}$ we find that, for any of the $\Delta M_R$ values explored, the results for the scattering length $a_1$ are of the order of one tenth or smaller than the experimental one, and the effective range becomes abnormally large, and we dismiss this case as a possible solution.

\section{Conclusions}
\label{sec:con}
We start from a $\Sigma^{\ast}$ state of non-molecular nature, which we would like to associate to the $\Sigma^*(1430)$ state, predicted theoretically within the chiral unitary approach, and recently discovered experimentally by the Belle Collaboration. The state has to couple to the $\pi \Lambda$ meson-baryon component, since it has been observed in this channel. Additionally, it will also couple to the channels coupled to $\pi \Lambda$ within a $SU(3)$ framework, namely $\bar K N$ and $\pi \Sigma$. We relate the coupling to these channels using a $SU(3)$ formalism. The unitary conditions in coupled channels then force the original state to couple to these components becoming a dressed state, with a non-molecular core and a meson-baryon cloud. We explore different scenarios, in which the bare state mass is close by or far away from the physical mass of the $\Sigma^*(1430)$ and find that if the bare state mass is very close by the $\Sigma^*(1430)$ mass, the state remains basically a non-molecular state. Yet, one pays a large price for it, since the width of the state is too small compared with experiment, and the effective range of the $\bar K^0 p$ channel becomes abnormally large. On the contrary, we find that a situation where the bare mass is about 100\,MeV higher than the $\Sigma^*(1430)$ mass does not show unacceptable results in the width and scattering parameters, but the state has become dressed by a large molecular cloud, very different in nature than the original state before its unitarization with the meson-baryon components. Future experiments determining the scattering parameters of the different coupled channels, available from measurements of the correlation functions, should provide elements to pin down the structure of that state more accurately. At present, the mass so close to threshold of the $\bar K N$ and the width of the state already measured, are sufficient to rule out the non-molecular nature of that state, but a more precise measurement of the $\bar K^0 p$ scattering length and the experimental determination of the $\bar K^0 p$ effective range should allow one to be more assertive in that statement. 

\section*{Acknowledgments}
This work is partly supported by the National Natural Science
Foundation of China under Grants  No. 12405089 and No. 12247108 and
the China Postdoctoral Science Foundation under Grant
No. 2022M720360 and No. 2022M720359. JX. Lin and J. Song wish to thank support from the China Scholarship Council. This work is also supported by
the Spanish Ministerio de Economia y Competitividad (MINECO) and European FEDER funds under
Contracts No. FIS2017-84038-C2-1-P B, PID2020-
112777GB-I00, and by Generalitat Valenciana under con-
tract PROMETEO/2020/023. This project has received
funding from the European Union Horizon 2020 research
and innovation programme under the program H2020-
INFRAIA-2018-1, grant agreement No. 824093 of the
STRONG-2020 project. This work is supported by the Spanish Ministerio de Ciencia e Innovaci\'on (MICINN) under contracts PID2020-112777GB-I00, PID2023-147458NB-C21 and CEX2023-001292-S; by Generalitat Valenciana under contracts PROMETEO/2020/023 and  CIPROM/2023/59.

\bibliographystyle{apsrev4-1}
\bibliography{ref}
\end{document}